# The Efficiency of Ideal Ferrielectric Energy Converter

**M.D. Zviadadze, I.G. Margvelashvili, A.G. Kvirikadze, L.A. Zamtaradze, A.I. Berdzenishvili**

I. Javakhishvili Tbilisi State University, 3 Chavchavadze Ave. Tbilisi, 0128, Georgia,

E. Andronikashvili Institute of Physics, 6 Tamarashvili Str., Tbilisi, 0177, Georgia

m.zviadadze@mail.ru

## Abstract

The article contains the calculation of the efficiency of ideal ferrielectric energy converter, the working body of which is ferrielectric and which converts the heat energy of low-potential water into the electric energy. Its operation is based on the sharp dependence of capacitivity of ferrielectric on the temperature of dielectric-ferrielectric structural phase transition in the vicinity of Curie temperature.

1. Recently, the interest to non-traditional energy sources has been sharply increased. The reason of such interest is the exhaustion of traditional fuels coal, oil and gas being imminent in the nearest future and the ecological implications caused by their increasing use.

The article contains the calculation of the efficiency of ideal ferrielectric energy converter (FEC), the working body (WB) of which is a capacitor with ferrielectric between the plates.

2. The principle of the operation of ferrielectric energy converter is absolutely identical to that of any heat engine. The use of ferrielectric as a working body is caused by the possibility of changing its capacitivity $\varepsilon(T)$ by $3 \div 6$ orders of magnitude in the region of room temperature at the change of temperature only by several degrees in the vicinity of Curie temperature $T_C$ [1-3].

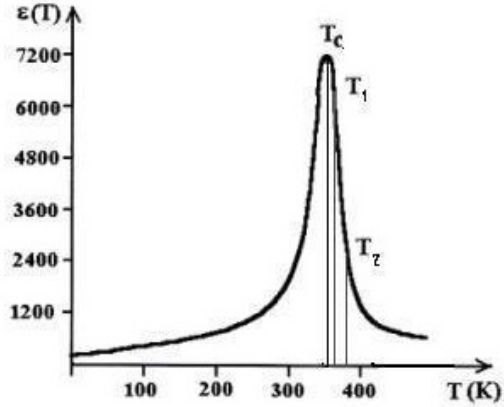

Fig. 1. Characteristic dependence of capacitivity of $BaTiO_3$ ferrielectric on temperature.

In FEC the heat energy of low-potential (LP) water is converted into electrostatic energy of ferrielectric capacitor, which can be used for doing a useful work. As in all heat engines, the work is performed at the expense of $\Delta Q = Q_H - Q_X$ difference, where $Q_H$ is the heat quantity received by WB from the heater and $Q_X$ is the heat quantity transferred by WB to the cooler. $\Delta Q$ difference is caused by the difference in physical conditions of heating and cooling of WB, which, in case of ferrielectric consists in the change of the strength $\vec{E}$ of uniaxial - одноосный electric field, in which the crystal is placed at heating and at cooling.

Electrostatic energy of the unit volume of dielectric in uniform electric field $\vec{E}$ is equal to $W_e = \vec{E} \cdot \vec{D}$, where $\vec{D} = \varepsilon_0 \, \varepsilon \, \vec{E}$ is the vector of electric induction.. In case of ferrielectric, the electrostatic energy of unit volume is determined by the expression $W_e = \varepsilon_0 \, \varepsilon(T) E^2$. At heating of ferrielectric from $T_1$ to $T_2$ under the conditions, when the uniform electric field in ferrielectric does not change, the internal energy of volume unit $U$ changes by the value

$$\Delta W = \varepsilon_0 [\varepsilon(T_2) - \varepsilon(T_1)] E^2 + \Delta W_l, \qquad (1)$$

where $\Delta W_l$ is the change of internal energy of crystal lattice of unit volume connected with the heat capacity of lattice $C_l(T)$ in the absence of electric field given by

$$\Delta W_l = \int_{T_1}^{T_2} C_l(T) dT = C_l(T_2 - T_1). \qquad (2)$$

Heat capacity of ferrielectric lattice in the absence of electric field $C_l(T)$ does not depend on the method of crystal heating, i.e. $C_{Pl} \approx C_{Vl}$. Approximation (2) is the consequence of the infinitesimal of electrostriction effect and of the infinitesimal of heat expansion coefficient, which for the unit volume is of the order of $10^{-6} \div 10^{-5} \, \text{deg.}^{-1}$. Besides, expression (2) assumes that in the region of considered temperatures the quantum corrections for the heat capacity are insignificant. As for the electrostatic heat capacity (the term in (1)) of much more complicated form, there is no necessity to know it, as the dependence of electrostatic energy on temperature. is known.

From the form of Eq. (1) it follows that in our approximation the ferrielectric is the sum of two independent subsystems – dipole and lattice. After the made remarks it is clear that the quantity of heat received by the crystal of V volume at its heating from $T_1$ to $T_2$ in the uniform electric field $\vec{E}$, according to the first law of thermodynamics, is equal to

$$Q_E = \left\{ \varepsilon_0 \left[ \varepsilon(T_2) - \varepsilon(T_1) \right] E^2 + C_l (T_2 - T_1) \right\} V \ . \tag{3}$$

For the quantity of heat $Q_0$ transferred by the crystal to the cooler at its cooling from $T_2$ to $T_1$ in the absence of electric field, the following expression is obtained:

$$Q_0 = C_l (T_2 - T_1) V \ . \tag{4}$$

3. As it is know well [2], the useful work performed by the system for one cycle, i.e. placing of the ferrielectric of $T_1$ temperature in the electric field $\vec{E}$ and its heating to the temperature $T_2$ with its following cooling to temperature $T_1$ in the absence of field, is determined both by the difference $Q_E - Q_0$, and by the conditions under which the cyclic process takes place.

For a detailed description of the mentioned cyclic process, let us consider a more real model, when the uniform electric field in ferrielectric is realized by a flat capacitor between the plates of which a ferrielectric is placed, and the heating and the cooling take place not at the constant $\vec{E}$, but at constant charge on plates $q$. In this case, the strength of electric field changes significantly due to the strong dependence of capacitivity on temperature. Thermodynamic cycle of FEC in variables $(q, U)$ is presented in Fig. 2.

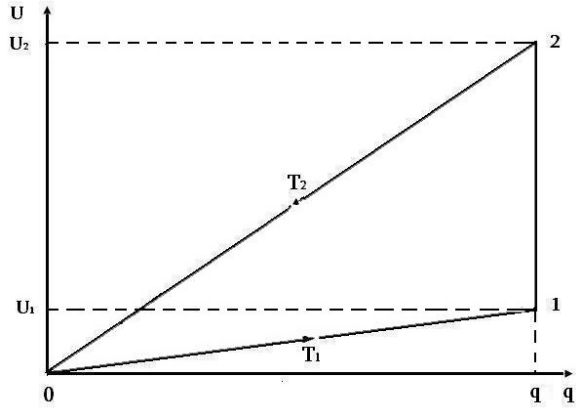

Fig. 2. Thermodynamic cycle of FEC $0 \rightarrow 1 \rightarrow 2 \rightarrow 0(T_2) \rightarrow 0(T_1)$ [2].

In this case, the charging (process $0 \rightarrow 1$) and discharging (process $2 \rightarrow 0$) of capacitor take place in isothermal regime, and the processes of heating $1 \rightarrow 2$ and cooling $0(T_2) \rightarrow 0(T_1)$ - in the electric field and at its absence, respectively.

The pattern of dependence $U(q)$ in the cycle $0 \rightarrow 1 \rightarrow 2 \rightarrow 0(T_2) \rightarrow 0(T_1)$ remains the same at charging and discharging of the capacitor in adiabatic regime due to the extreme infinitesimal of electrocaloric effect.

Further, it is assumed (as it is shown in Fig. 1) that the initial and the final temperatures $T_1$ and $T_2$ lie more to the right of Curie temperature, i.e. $T_C < T_1 < T_2$ and the condition $T_1 \approx T_C$ is fulfilled. Such choice of temperature allows to avoid the difficulties connected with the dependence of $\varepsilon(T)$ from electric field

The quantity of heat received by ferrielectric capacitor at increasing its temperature from $T_1$ to $T_2$ without changing $Q$ charge equals

$$Q_q = E(T_1)(K-1) + C_l V(T_2 - T_1), \qquad (5)$$

where the following notations are inserted

$$K = C(T_1)/C(T_2) = \varepsilon(T_1)/\varepsilon(T_2) > 1, \ E(T_1) = q^2/(2C(T_1)), \ C(T) = \varepsilon_0 \varepsilon(T) S/d . \qquad (6)$$

The quantity of heat $Q_0$, transferred by the discharged capacitor to the cooler, is determined by expression (4), and the difference

$$Q_q - Q_0 = A_C = E(T_1)(K-1) \qquad (7)$$

determines the work $A_C$ during a cycle. For charging the capacitor the work $A_q = E(T_1)$ is spent, therefore, the useful work done by the subsystem of electric dipoles equals

$$A = A_C - A_q = E(T_1)(K-2) \qquad (8)$$

and the efficiency will be

$$\eta = \frac{A}{Q_q} = \frac{K-2}{K-1} \cdot \frac{1}{1 + \dfrac{2C_l V}{E(T_1)} \dfrac{T_2 - T_1}{K-1}} . \qquad (9)$$

In experiment the condition $K >> 1$ is easy to obtain. In this case, for the efficiency of FEC we have the estimate

$$\eta = \left[ 1 + \frac{2C_l V}{E(T_2)} (T_2 - T_1) \right]^{-1} . \qquad (10)$$

**Acknowledgements:** This work is supported by the Shota Rustaveli National Science Foundation, Grant **GNSF 712/07**. The authors are grateful to M. Nikoladze for her help in preparing the article.